\documentclass[letter,twocolumn, nofootinbib,showkeys,showpacs,10pt]{revtex4}
\usepackage{amssymb}
\usepackage{amsmath}
\usepackage{makeidx}
\usepackage{amsfonts}
\usepackage{graphicx,epstopdf}
\usepackage[ansinew]{inputenc}
\usepackage[usenames,dvipsnames]{pstricks}
\usepackage{subfigure}
\usepackage{pdfpages}
\usepackage{epsfig}
\usepackage{pst-grad}
\usepackage{pst-plot}
\usepackage[colorlinks,hyperindex]{hyperref}

\hypersetup{
colorlinks,
citecolor=blue,
linkcolor=black,
urlcolor=black,
}

\newcommand{\be}{\begin{equation}}
\newcommand{\ee}{\end{equation}}

\newcommand{\beq}{\begin{eqnarray}}
\newcommand{\eeq}{\end{eqnarray}}

\begin{document}

\title{D-Oscillons in the Standard Model-Extension }
\author{R. A. C. Correa}
\email{rafael.couceiro@ufabc.edu.br}
\affiliation{  CCNH, Universidade Federal do ABC, 09210-580, Santo André, SP, Brazil}
\author{Rold\~ao da Rocha}
\email{roldao.rocha@ufabc.edu.br}
\affiliation{CMCC, Universidade Federal do ABC, 09210-580, Santo André, SP, Brazil}
\affiliation{International School for Advanced Studies (SISSA), Via Bonomea 265, 34136
Trieste, Italy}
\author{A. de Souza Dutra}
\email{dutra@feg.unesp.br}
\affiliation{Universidade Estadual Paulista (UNESP), Campus de Guaratinguetá, 12516-410,
SP, Brazil}

\begin{abstract}
In this work we investigate the consequences of the Lorentz symmetry
violation on extremely long-living, time-dependent, and spatially localized
field configurations, named oscillons. This is accomplished  for two interacting scalar field theories in ($D+1$)
dimensions in the context of the so-called
Standard Model-Extension. We show that $D$-dimensional scalar field
lumps can present a typical size $R_{\min }\ll R_{KK}$, where $R_{KK}$ is
the extent of extra dimensions in Kaluza-Klein theories.
The size $R_{\min }$ is shown to strongly depend upon the terms that
control the Lorentz violation of the theory. This implies either contraction
or dilation of the average radius $R_{\min}$, and a new rule for its
composition, likewise. Moreover, we show that the spatial dimensions for
existence of oscillating lumps have an upper limit, opening new
possibilities to probe the existence of $D$-dimensional oscillons at TeV
energy scale. In addition, in a cosmological scenario with Lorentz symmetry
breaking, we show that in the early Universe with an extremely high energy
density and a strong Lorentz violation, the typical size $R_{\min }$ was
highly dilated. As the Universe had expanded and cooled down,
it then passed through a phase transition towards a Lorentz
symmetry, wherein $R_{\min}$ tends to be compact.
\end{abstract}

\maketitle

\section{Introduction}

The Lorentz invariance represents the essential symmetry in the Standard
Model of elementary particles. Notwithstanding, Lorentz symmetry may be violated at high energies \cite{kostelecky1}, constituting thus a
fundamental tool in several fields \cite%
{kosteleccky4,kostelecky5,Bernardini:2008ef,Bernardini:2007ez,Bernardini:2007ex,musad}%
. For instance, by using a scalar-vector-tensor theory with Lorentz
violation, the exact Lorentz violation inflationary solutions can be found without an inflaton potential \cite{kanno}. Topological defects in
a Lorentz symmetry violation (LSV) framework have been recently addressed 
\cite{dutra1,menezes,rafael1}, providing in particular the LSV as an
asymmetry between defects and anti-defects  \cite{barreto}.
Motivated by these results, travelling solitons in Lorentz and \textit{CPT}
breaking systems were studied \cite{rafael1}, where the solutions present a
critical behavior controlled by the choice of a scalar. Other prominent
interests regarding LSV further arise in various contexts, encompassing, e.
g., gravity, monopoles and vortices \cite%
{manoel2,julio,Ferrari:2006gs}.

Topologically stable configurations play a prominent role on non-linear
models. Among non-linear field configurations, a distinguished class of
time-dependent stable solutions are exemplified by the breathers, in sine-Gordon like models. Another time-dependent field
configuration whose stability is granted by charge conservation are the
so-called $Q$-balls, as named by Coleman \cite{S.Coleman1}, or alternatively
non-topological solitons \cite{Lee}. However, considering the fact that many
physical systems interestingly may present a metastable behavior, a further
class of non-linear systems may present a very long-living configuration,
usually known as oscillon. This class of solutions was discovered by
Bogolyubsky and Makhankov \cite{I.L.Bogolyubsky}, and then rediscovered
posteriorly by Gleiser \cite{Gleiser1}. These solutions appeared in the
study of the dynamics of first-order phase transitions and bubble
nucleation. Since then, an increasing amount of works has been dedicated to
the study of these objects \cite%
{Gleiser1,Gleiser2,g1,g2,Gleiser4,Honda,Gleiser5,Farhi,Graham1,Graham2,Linde,Gleiserintj,Kolb,P. M. Saffin,Guth,H. Arodz,Gyula1,Gleiser8,Gleiser9,Gleiser11,Gyula2,Hertzberg,Mustafa,AHEP-RAFAEL-ALVARO,Gleiser10,Segur1,PRL105(2010)}%
.

Oscillons are quite general configurations found in various contexts, as the
Abelian-Higgs $U(1)$ models \cite{Gleiser8}, the standard model $SU(2)\times
U(1)$ \cite{Farhi, Graham2}, inflationary cosmological models \cite%
{Linde,Gleiserintj}, axion models \cite{Kolb}, expanding Universe scenarios 
\cite{Graham1, Guth, Mustafa} and systems involving phase transitions as
well \cite{Gleiser2}. In a recent work by Gleiser \textit{et al.} \cite%
{Gleiser11} the problem of the hybrid inflation characterized by two real
scalar fields interacting quadratically was analyzed, where a new class of
oscillons arises both in excited and ground states as well.

The usual oscillon aspect is typically that of a bell shape which oscillates
sinusoidally. They are long lived time-dependent $D$-dimensional scalar
field lumps. Moreover, their lifetimes are long enough to yield noteworthy
effects. In fact, their collapse happens in very short time scales, thus
they might be evinced in large compact extra dimensions frameworks as an
abrupt burst of particles from a small region \cite{Gleiser5}. Recently,
Amin and Shirokoff have shown that depending upon the intensity of the
self-interacting scalar field coupling constant, it is possible to observe
oscillons with a kind of plateau at its top \cite{Mustafa}. Indeed, these
new oscillons were shown to be more robust with respect to collapse
instabilities in three spatial dimensions. In a recent work, the impact of
the Lorentz and CPT breaking symmetries was discussed in the context of the
so-called flat-top oscillons \cite{AHEP-RAFAEL-ALVARO}.

At this point it is worth to remark that Segur and Kruskal \cite{Segur1}
have shown that the asymptotic expansion does not represent in general an
exact solution for the scalar field. Indeed, it simply represents a first
order asymptotic expansion. They also showed in one spatial dimension that the
solutions radiate as well \cite{Segur1}. Besides, the computation of
the emitted radiation of the oscillons was extended for two and three
spatial dimensions \cite{Gyula2}. Another important result was put forward
by Hertzberg \cite{Hertzberg}, computing the decaying rate of quantized
oscillons and showing moreover that the quantum rate decay is very distinct
from the classical one.

Our main aim here is to show that in a LSV framework oscillons present a
typical size $R_{\min}$ that strongly depends upon the LSV parameters. In
addition we shall prove that in the early Universe, with an extremely high
energy density and strong Lorentz violation, the typical size $R_{\min }$ is
characterized by a high dilation, in a cosmological scenario with LSV. We
show that as the Universe expanded and cooled down, {\color{black}{occurring}%
} a phase transition towards a Lorentz symmetry framework, the oscillon size 
$R_{\min } $ is compacted.

This work is organized as follows: in Sect. II we introduce a Lagrangian
regarding the Standard Model-Extension. Hence, a two scalar fields theory,
in ($D+1$)-dimensional space-time, is taken into account in a LSV framework.
In Sect. III oscillons configurations are thus modelled with respect to a
Gaussian formulation. Thus in Sect. IV the stability of the oscillons in the
Standard Model-Extension is studied and some kinds of potentials, including
the quadratic, cubic and quartic ones, are analyzed, providing a minimal
size for the oscillon. Moreover, we show that the dimension for the
existence of oscillating lumps presents an upper limit. Sect. VI provides an
application of the studied framework, concerning a double-well potential.

\bigskip

\section{Standard Model-Extension Lagrangian in $D$ dimensions}

In this section the Lagrangian in the Standard Model-Extension (SME) context
is introduced, describing a theory with two scalar fields in ($D+1$%
)-dimensional space-time in a Lorentz symmetry breaking framework. Our aim
in working with a two scalar fields theory with Lorentz violation (LV) comes
from the fact that such theories can provide observable effects. Therefore,
the approach is much more favorable from an experimental point of view. 
{Here, we will consider a local field theory in which the Lagrangian }%
$L$ {can be written as a volume integral over a density function }$%
\mathcal{L}$%
\begin{equation}
L=\int d^{D}x\;\mathcal{L}(\phi ,\chi ,\dot{\phi},\dot{\chi},\mathbf{\nabla }%
\phi ,\mathbf{\nabla }\chi ),  \label{L1}
\end{equation}
\noindent {where the Lagrangian density may depend on the field
functions }$\phi (\mathbf{x},t)$ {and }$\chi (\mathbf{x},t)${%
, on its time derivatives }$\dot{\phi}(\mathbf{x},t)${}{\ and }$\dot{%
\chi}(\mathbf{x},t)${, and also on the gradients }$\mathbf{\nabla }%
\phi (\mathbf{x},t)${}{\ and }$\mathbf{\nabla }\chi (\mathbf{x},t)$%
{}{. It is important to remark that the restriction to
local Lagrange density is sufficiently general for the framework of all recent field theories.}

{}{\ Furthermore, }since any deformation away from spherical symmetry
leads to more energetic configurations \cite{S.Coleman1}, {}{we will
deal with} spherically symmetric field configurations.{}{\ In this
case, we can write the }$D${}{-dimensional spherical coordinate system
as}%
\begin{eqnarray}
x_{1} &=&r\sin \theta _{1}\cdots \sin \theta _{D-1},  \notag \\
x_{2} &=&r\sin \theta _{1}\cdots \sin \theta _{D-2}\cos \theta _{D-1}, 
\notag \\
&\vdots& \\
x_{k} &=&r\sin \theta _{1}\cdots \sin \theta _{D-k}\cos \theta _{D-k+1},%
\text{ \ }2\leq k\leq D-1,  \notag \\
&\vdots&  \notag \\
x_{D} &=&r\cos \theta _{1}.  \notag
\end{eqnarray}

{}{Here, it is important to highlight that this system is the
generalization of spherical coordinates in three dimensions.} {}{Thus,
using the above spherical coordinates system yields the volume element} $d^{D}x=r^{D-1}drd\Omega _{D},$ {}{where }$d\Omega _{D}$ {}{is the so-called element of
the }$D${}{-dimensional solid angle, given by}%
\begin{equation}
\!\!\!\!d\Omega _{D}=\sin ^{D-2}\theta _{1}\sin ^{D-3}\theta _{2}\cdots \sin \theta
_{D-2}d\theta _{1}\cdots d\theta _{D-1},
\end{equation}

{}{Now, the integral }%
\begin{equation}
\int\limits_{0}^{\pi }d\theta \sin ^{n}\theta =\frac{\sqrt{\pi }\Gamma
\lbrack (n+1)/2]}{\Gamma \lbrack (n+2)/2]},
\end{equation}%
{}{ yields  the total solid
angle in }$D$ {}{dimensions} $
\Omega _{D}=\frac{2\pi ^{D/2}}{\Gamma (D/2)}.$

{}{Thus, we can rewrite the Lagrangian (\ref{L1}), in }$D${}{%
-dimensional spherical coordinates}, {}{in the form}%
\begin{equation}
L=\frac{2\pi ^{D/2}}{\Gamma (D/2)}\int drr^{D-1}\mathcal{L}(\phi ,\chi ,\dot{%
\phi},\dot{\chi},\mathbf{\nabla }\phi ,\mathbf{\nabla }\chi ),  \label{L2}
\end{equation}

\noindent {}{where the fields }$\phi ${}{\ and }$\chi ${}{\
now} {}{are functions of }$r=\sqrt{x_{1}^{2}+\cdots +x_{D}^{2}}$, 
{}{and also of }$t${}{.}

{}{However, in order to work with a LSV theory, 
we assume that the Lagrangian density }$L${}{\ has the form}
\begin{equation}
\mathcal{L}=\frac{1}{2}\partial _{a}\phi \partial ^{a}\phi +\frac{1}{2}%
\partial _{a}\chi \partial ^{a}\chi +K^{ab}\partial _{a}\phi \partial
_{b}\chi -V(\phi ,\chi ),  \label{1.11}
\end{equation}%
\noindent {}{where }$a,b=0,1${}{\ and }$V(\phi ,\chi )${}{\
denotes the scalar field potential.} {}{Also, we are using the
following definition} 
\begin{equation}
\partial _{a}:=(\partial /\partial t,\partial /\partial r),\text{ }\partial
^{a}:=(\partial /\partial t,-\partial /\partial r).
\end{equation}

Moreover, in the above Lagrangian (\ref{1.11})%
\begin{equation}
K^{ab}={\small {\left( 
\begin{array}{cc}
K^{00} & K^{01} \\ 
K^{10} & K^{11}%
\end{array}%
\right) ,}}
\end{equation}%
is a dimensionless rank-2 symmetric tensor that encompasses the Lorentz
symmetry breaking. It is worth to emphasize that in general $K^{ab}$ has
arbitrary parameters, but if this matrix is real, symmetric, and traceless,
the \textit{CPT }symmetry is conserved \cite{k1, k2, k3, k4}. Moreover,
under \textit{CPT}{}{\ operation}, namely $\partial _{a}\mapsto
-\partial _{a}$, t{}{he term} $K^{ab}\partial _{a}\phi \partial
_{b}\chi $ {}{goes as} $K^{ab}\partial _{a}\phi \partial _{b}\chi
\mapsto +K^{ab}\partial _{a}\phi \partial _{b}\chi $. Thus, $K^{ab}$ is 
\textit{CPT}-even \cite{AHEP-RAFAEL-ALVARO,k4}. Furthermore, the tensor $%
K^{ab}$ should be symmetric in order to avoid a vanishing contribution. The
LSV parameters are denoted by $K^{00}\sim K^{11}=\alpha $ and $K^{01}\sim
K^{10}=\beta $.

It is appropriate to stress that in (\ref{1.11}) the coefficients for LV
cannot be removed from the Lagrangian by using either variable or field
redefinitions. In fact, coordinate choices and field redefinitions solely
make the Lorentz symmetry violation to go to another sector of the theory.

The aim here is to analyze the behavior of oscillons in a class of
potentials that are as general as possible. We then consider systems that
can be decoupled by applying a field redefinition 
\begin{equation}
\phi =\frac{\psi +\sigma }{\sqrt{2}},\qquad \chi =\frac{\psi -\sigma }{\sqrt{%
2}}\,.  \label{1.3}
\end{equation}%
Hence, the preceding rotation shows that it is possible to put the
Lagrangian (\ref{L2}) in the form%
\begin{eqnarray}
L\!\! &=&\!\!\!\left. \frac{2\pi ^{D/2}}{\Gamma (D/2)}\!\int \!\!drr^{D-1}%
\left[ \frac{1}{2}(1\!+\!\alpha )(\partial _{t}\psi )^{2}+\frac{1}{2}%
(1\!-\!\alpha )(\partial _{t}\sigma )^{2}\right. \right.  \notag \\
&&  \notag \\
&&\!\!\left. -\frac{1}{2}(1\!-\!\alpha )(\partial _{r}\psi )^{2}-\frac{1}{2}%
(1\!+\!\alpha )(\partial _{r}\sigma )^{2}+\beta \partial _{t}\psi \,\partial
_{r}\psi \right.  \notag \\
&&  \notag \\
&&\left. \left. -\beta \partial _{t}\sigma \partial _{r}\sigma -V(\psi
,\sigma )\right] .\right.  \label{1.41}
\end{eqnarray}

Let us now consider a class of potentials {}{\ }$V(\phi ,\chi )${}{%
\ }such that the rotation enables to write 
\begin{equation}
V(\psi ,\sigma ):=U_{1}(\psi )+U_{2}(\sigma )\,,  \label{1.5}
\end{equation}
where $U_{1}$ and $U_{2}$ are arbitrary. In this case, we can find the
original potential described in terms of the fields $\phi$ and $\chi$,
performing the rotation (\ref{1.3}) back.

Note that under the condition (\ref{1.5}) the Lagrangian reads a sum of two
independent Lagrangians 
\begin{equation*}
L=\sum_{j=1}^{2}L_{j},
\end{equation*}%
\noindent with%
\begin{eqnarray}
&&\left. L_{j}=\frac{2\pi ^{D/2}}{\Gamma (D/2)}\int drr^{D-1}\left[
A_{j}(\partial _{t}\Phi _{j})^{2}-B_{j}(\partial _{r}\Phi _{j})^{2}\right.
\right.  \notag \\
&&  \notag \\
&&\left. \left. \qquad \qquad C_{j}(\partial _{t}\Phi _{j})(\partial
_{r}\Phi _{j})-U_{j}(\Phi _{j})\right] \,,\right.  \label{1.71}
\end{eqnarray}%
\noindent where the following notation 
\begin{eqnarray}
\Phi _{1} &=&\psi ,\qquad \quad \text{ }\Phi _{2}=\sigma ,  \notag \\
A_{j} &=&[1+(-1)^{j+1}\alpha ]/2,  \label{aj} \\
B_{j} &=&[1+(-1)^{j}\alpha ]/2,  \label{bj} \\
C_{j} &=&(-1)^{j+1}\beta \,,  \label{cj}
\end{eqnarray}%
\noindent is used.

It is important to remark that the Lagrangian (\ref{1.71}) is represented by
a sum of two independent Lagrangians, where the fields $\phi $ and $\chi $
are connected through a inverse rotation with respect to Eq. (\ref{1.3}). In
the next section, we show the analytic approach to find oscillons.

\section{Modelling Oscillons Configurations}

The analytic approach to investigate the dynamics of time-dependent
oscillating scalar field configurations was introduced by Gleiser \cite{Gleiser5}. Motivated by
numerical investigations \cite{Gleiser1,Gleiser4}, Gleiser showed that
oscillons are well approximated by a Gaussian curve. Thus, in the present
work, we also assume that oscillons solutions can be modelled as%
\begin{equation}
\Phi _{j}(r,t)=G_{j}(t)\exp \left( \frac{-q_{j}r^{2}}{R_{j}^{2}}\right)
+\Phi _{j}^{v}.  \label{1.8}
\end{equation}

\noindent where $q_{j}>0$ and $\Phi _{j}^{v}$ is the asymptotic value when $
r\rightarrow \infty $, which is determined by the form of the potentials $
U_{j}(\Phi _{j})$. {}{At this point, it is important to remark that any ansatz  is an assumption, where boundary conditions can be taken
into account. After an ansatz  has been established, the equations are solved
for the general function. Hence we can verify the validity of the assumption
provided by the ansatz. Eq. (\ref{1.8}) is the natural choice. In fact, as
shown in Refs. \cite{Gleiser1,Gleiser2}, oscillons can be found by  profiles either Gaussian or hyperbolic tangent type. Here we use the Gaussian ansatz  in Eq. (\ref{1.8}), where %
$G_{j}(t)$ is an amplitude that can be identified with }$\Phi
_{j}(0,t)-\Phi _{j}^{v}$, where $R_{j}${}{\ denotes the core
radius. Similarly to the case of Lorentz symmetry, our case regarding
LSV violation makes the evolution of the configuration to split
into different stages. The lifetime of the oscillon configuration 
is sensitive to the choices of }$R_{j}${}{\ and }$\Phi _{j}(0,t)$. 
{}{The lifetime of the oscillon configuration is thus related to
perturbations arising by the different choices of initial parameters, what
increases the amount of radiation being emitted. As shown in Ref. \cite%
{Gleiser1}, due to the tiny, however steady,  oscillon radiation, at some
point the maximum amplitude decreases below the inflection point of the
potential and }$\Phi _{j}(0,t)\rightarrow \Phi _{j}^{v}${}{\
exponentially fast. During the nonlinear evolution of these configurations a
regime of dynamical stability was shown to be achieved, where the energy
is conserved within a localized region} {}{\ \cite%
{Gleiser1,Gleiser2,g1,g2}. Such a Gaussian ansatz  is further acquired by
numerical analysis in an overlapping context. The analytical results
obtained were verified numerically to great accuracy, confirming to be
adequate for the goals investigated in, e .g., \cite{Gleiser1,Gleiser2,g1,g2}. 
}

{}{Here we are interested in the particular case of polynomial
potentials, which are written as} 
\begin{equation}
U_{j}(\Phi _{j})=\sum\limits_{n=0}^{N}\frac{g_{j,n}}{n!}\Phi
_{j}^{n}-U_{j}(\Phi _{j}^{v}),  \label{1.9}
\end{equation}%
\noindent where $N$ denotes the maximal power of the potential, $g_{j,n}$ 
are scalars, and the vacuum energy $U_{j}(\Phi _{j}^{v})$ is taken away from
the potential to prevent spurious divergences upon spatial integration. An
immediate consequence of the definition of $U_{j}(\Phi _{j})$ in Eq. (\ref%
{1.9}) is that the potential $V(\Phi _{1},\Phi _{2})$ now takes the form 
\begin{equation}
V(\Phi _{1},\Phi _{2})=\sum\limits_{n=0}^{N}\frac{g_{1,n}}{n!}\Phi
_{1}^{n}+\sum\limits_{n=0}^{N}\frac{g_{2,n}}{n!}\Phi _{2}^{n}-U(\Phi
_{1}^{v},\Phi _{2}^{v}),  \label{1.10}
\end{equation}%
\noindent where $U(\Phi _{1}^{v},\Phi _{2}^{v})=U_{1}(\Phi
_{1}^{v})+U_{2}(\Phi _{2}^{v})$ is the vacuum energy.

Hence, using Eqs. (\ref{1.8}) and (\ref{1.9}) into Eq. (\ref{1.71}), and
integrating over space coordinates, it yields 
\begin{eqnarray}
&&\left. L_{j}=\pi ^{D/2}\left\{ \varepsilon _{j}^{-D/2}A_{j}\dot{G}_{j}^{2}-%
\frac{2q_{j}^{2}}{R_{j}^{4}}{D\varepsilon _{j}^{-(D+2)/2}B_{j}G_{j}^{2}}%
\right. \right.  \notag \\
&&  \notag \\
&&\left. -\frac{q_{j}}{R_{j}^{2}}\,\varepsilon _{j}^{-(D+1)/2}\frac{\Gamma
\lbrack (D+1)/2]}{\Gamma (D/2)}C_{j}G_{j}\dot{G}_{j}\right.  \notag \\
&&  \notag \\
&&\left. \left. -\sum_{n=2}^{N}\frac{G_{j}^{n}}{n!}\varepsilon
_{j,n}^{-D/2}U_{j}^{(n)}(\Phi _{j}^{v})\right\} ,\right.  \label{2.11}
\end{eqnarray}%
\noindent where $\varepsilon _{j}:=2q_{j}/R_{j}^{2}$, $\varepsilon
_{j,n}:=nq_{j}/R_{j}^{2}$, $U_{j}^{(n)}(\Phi _{j}^{v}):=dU_{j}^{n}(\Phi
_{j}^{v})/d\Phi ^{n}$, and the dot stands for the derivative with respect to
time.

\section{Stability of the Oscillons in the Standard Model-Extension}

In the preceding sections we introduced the Lagrangian{}{\ }in the SME
context. In this section, we are going to deduce the equations of motion for
the functions $G_{j}(t)$ in (\ref{1.8}) and introduce the so-called
effective frequency $\omega _{j}$ as well, which is necessary to examine the
stability of the oscillon. Hence, from the Lagrangian (\ref{2.11}) the
corresponding equations of motion read 
\begin{eqnarray}
&&\left. \ddot{G}_{j}+\frac{2Dq_{j}^2}{R_{j}^{4}} \frac{B_{j}}{%
A_{j}\varepsilon _{j}}G_{j}\right.  \notag \\
&&  \notag \\
&&\left. +\frac{\varepsilon _{j}^{D/2}}{2A_{j}}\sum_{n=2}^{N}\frac{G^{n-1}}{%
(n-1)!}\varepsilon _{j,n}^{-D/2}U_{j}^{(n)}(\Phi _{j}^{v})=0.\right.
\label{3.1}
\end{eqnarray}

In order to analyze the stability of the system, let us expand the amplitude
as 
\begin{equation}
G_{j}(t)=G_{j,0}(t)+\delta G_{j}(t),  \label{3.2}
\end{equation}
\noindent where $G_{j,0}(t)$ is the solution of Eq. (\ref{3.1}), and $\delta
G_{j}(t)$ describes a small perturbation.

Therefore, by applying Eq. (\ref{3.2}) into Eq. (\ref{3.1}) and subsequently
linearizing the result, it yields 
\begin{eqnarray}
&&\left. \delta \ddot{G}_{j}=-\frac{DB_{j}}{2A_{j}\varepsilon _{j}}\left( 
\frac{2q_{j}}{R_{j}^{2}}\right) ^{2}\right.  \notag \\
&&  \notag \\
&&\left. +\frac{\varepsilon _{j}^{D/2}}{2A_{j}}\left[ \sum_{n=2}^{N}\frac{%
G_{j,0}^{n-2}}{(n-2)!}\varepsilon _{j,n}^{-D/2}U_{j}^{(n)}(\Phi _{j}^{v})%
\right] \delta G_{j}.\right.  \label{3.3}
\end{eqnarray}

In the light of the analysis accomplished in this section, the following
effective frequency is introduced 
\begin{eqnarray}
&&\left. \omega _{j}^{2}(R_{j},A_{j},B_{j},G_{j,0}):=\frac{DB_{j}}{%
2A_{j}\varepsilon _{j}}\left( \frac{2q_{j}}{R_{j}^{2}}\right) ^{2}\right. 
\notag \\
&&  \notag \\
&&\left. +\frac{\varepsilon _{j}^{D/2}}{2A_{j}}\left[ \sum_{n=2}^{N}\frac{%
G_{j,0}^{n-2}}{(n-2)!}\varepsilon _{j,n}^{-D/2}U_{j}^{(n)}(\Phi _{j}^{v})%
\right] .\right.  \label{3.4}
\end{eqnarray}

We can see, from the above effective frequency, that this quantity leads to
a distinct definition of frequency in a scenario with LSV. Indeed, here
there is an explicit dependence of the parameters responsible by the Lorentz
violation.

For simplicity, let us use the above definition to rewrite Eq.(\ref{3.3}) in
the form 
\begin{equation}
\delta \ddot{G}_{j}=-\omega _{j}^{2}(R_{j},A_{j},B_{j},G_{j,0})G_{j}.
\label{3.5}
\end{equation}
The above equation enables us to understand the stability of the motion. In
fact, if $\omega _{j}^{2}<0$, instabilities occur. On the other hand, when $%
\omega _{j}^{2}>0$, the system is stable.

As a straightforward example, consider the case where $U_{j}(\Phi _{j})=0$.
Consequently the effective frequency becomes%
\begin{equation}
\omega _{j}^{2}=\frac{D}{R_{j}^{2}}\left( \frac{q_{j}B_{j}}{A_{j}}\right) .
\label{3.6}
\end{equation}
The above established result prominently enables us to show that the
frequency is a function of the parameters responsible by the effects of the
Lorentz violation.

\subsection{Quadratic potentials}

In this section we will consider the case of quadratic potentials, where $%
U_{j}(\Phi _{j})$ is given by%
\begin{equation}
U_{j}(\Phi _{j})= g_{j,1}\Phi _{j}+\frac{g_{j,2}\Phi _{j}^{2}}{2}-U(\Phi
_{j}^{v}).  \label{3.7}
\end{equation}

In this case, we have%
\begin{equation}
\omega _{j}^{2}=\frac{D}{R_{j}^{2}}\left( \frac{q_{j}B_{j}}{A_{j}}\right) +%
\frac{U^{(2)}(\Phi _{j}^{v})}{2A_{j}}.  \label{3.8}
\end{equation}

Therefore, imposing that $A_{j}, B_{j}$, and $U^{(2)}(\Phi _{j}^{v})$ are
positive it implies that $\omega _{j}^{2}>0$, precluding thus any kind of
instability. On the other hand, if $A_{j}>0$, $B_{j}>0$, and $U^{(2)}(\Phi
_{j}^{v})<0$, instabilities are possible when%
\begin{equation}
\frac{D}{R_{j}^{2}}<\frac{\left\vert U^{(2)}(\Phi _{j}^{v})\right\vert }{%
2q_{j}B_{j}},  \label{3.81}
\end{equation}
\noindent which provides a minimal values for the oscillon typical size: 
\begin{equation}
R_{j}\geq \sqrt{2q_{j}B_{j}}\left[ \frac{D}{\left\vert U^{(2)}(\Phi
_{j}^{v})\right\vert }\right] ^{1/2}.  \label{3.82}
\end{equation}

Nevertheless, instabilities are allowed when $A_{j}>0$, $B_{j}<0$ and $%
U^{(2)}(\Phi _{j}^{v})<0$, such that%
\begin{equation}
R_{j}\geq \sqrt{2q_{j}\left\vert B_{j}\right\vert }\left[ \frac{D}{%
U^{(2)}(\Phi _{j}^{v})}\right] ^{1/2}.  \label{3.83}
\end{equation}

At this point, we observe that once one recover the expression of the
original fields $\phi $ and $\chi $ by using the results obtained for $\Phi
_{1}$ and $\Phi _{2}$, the resulting minimal size of the $\phi $ and $\chi $
oscillons is approximately the one of the biggest of the decoupled fields $%
\Phi _{1}$ and $\Phi _{2}$. This can seen in the case plotted in the Figure
1. In fact, this is a general feature of these configurations and will
equally appear in the next examples.

\subsection{Cubic potentials}

Now, in this section we will analyze the case of the cubic potential.
Therefore, we consider the following potential%
\begin{equation}
U_{j}(\Phi _{j})=g_{j,1}\Phi _{j}+\frac{g_{j,2}\Phi _{j}^{2}}{2!}+\frac{%
g_{j,3}\Phi _{j}^{3}}{3!}-U(\Phi _{j}^{v}).  \label{3.9}
\end{equation}

At this point, it is important to remark that in the cubic potential the
parity is broken. Consequently, the potential has an inflection point which
is determined by the relation $U_{j}^{(2)}(\Phi _{j})=0$. As a consequence,
the inflection point is provided by 
\begin{equation}
\Phi _{j}^{\inf}=-g_{j,2}/g_{j,3}.
\end{equation}

For simplicity, however without loss of generality, we take $g_{j,1}=0$.
Thus, the vacuum state corresponds to the values 
\begin{eqnarray}
\Phi _{j}^{v} = 
\begin{cases}
0, & \text{ for }g_{j,2}>0, \\ 
-2g_{j,2}/g_{j,3}, & \text{ for }g_{j,2}<0.%
\end{cases}%
\end{eqnarray}

Now, by using the potential (\ref{3.9}) we can show that the effective
frequency can be written as 
\begin{equation}
\hspace*{-0.2cm}\omega _{j}^{2}=\!\frac{D}{R_{j}^{2}}\!\left(\frac{q_{j}B_{j}%
}{A_{j}}\right) \!+\! \frac{U^{(2)}(\Phi _{j}^{v})}{2A_{j}}\!+\!\left( \frac{%
2}{3}\right)^{D/2}\!\!\frac{G_{j,0}U^{(3)}(\Phi _{j}^{v})}{2A_{j}},  \notag
\end{equation}
\noindent which reads 
\begin{equation}
\omega _{j}^{2}=\frac{D}{R_{j}^{2}}\left( \frac{q_{j}B_{j}}{A_{j}}\right) +%
\frac{g_{j,3}}{2A_{j}}\left[ \frac{g_{j,2}}{g_{j,3}}+\Phi _{j}^{v}+\left( 
\frac{2}{3}\right) ^{D/2}G_{j,0}\right] .  \notag
\end{equation}

To satisfy the condition $\omega _{j}^{2}<0$, which is necessary to
guarantee the existence of oscillons, the sign of $G_{j,0}$ must be opposite
to that of $g_{j,3}$. Indeed, long-lived oscillons can exist if the
oscillations above the vacuum of $U_{j}^{(2)}(\Phi _{j})<0$ for a sustained
period of time. Therefore, the relations of existence, to be analyzed in the
following sub-subsections, hold. In addition, in the following cases the
minimal radius explicitly depends of the LSV parameter $\alpha $, through
the $B_{j}$ in Eq.(\ref{bj}).

\subsubsection{For $g_{j,2}>0$ and $g_{j,3}\gtrless 0$}

Here, we find that%
\begin{equation}
R_{j}\geq \sqrt{\frac{2Dq_{j}B_{j}}{\left\vert g_{j,3}\right\vert \left[
-\Phi _{j}^{\inf }+\left( \frac{2}{3}\right) ^{D/2}G_{j,0}\right] }}.
\label{3.112}
\end{equation}

\subsubsection{For $g_{j,2}<0$ and $g_{j,3}\gtrless0$}

In this case, it follows that the oscillon radius obeys the constraint 
\begin{equation}
R_{j}\geq \sqrt{\frac{2Dq_{j}B_{j}}{\left\vert g_{j,3}\right\vert \left[
\Phi _{j}^{\inf }+\left( \frac{2}{3}\right) ^{D/2}\left\vert
G_{j,0}\right\vert\right] }}.  \label{3.114}
\end{equation}

Moreover, since that $R_{j}^{2}$ must be a positive number, the amplitudes $%
G_{j,0}$ must obey the condition%
\begin{equation*}
\left\vert G_{j,0}\right\vert \lesseqqgtr \left( \frac{3}{2}\right)
^{D/2}\left\vert \Phi _{j}^{\inf }\right\vert \text{, \quad for }%
B_{j}\lessgtr 0\,.
\end{equation*}

Again, as advertised in the previous section, the minimal radius of the $%
\phi $ and $\chi $ fields will be the bigger one between $R_{1}$ and $R_{2}$.

\subsection{Quartic potentials}

We are now going to study the case of quadratic potentials, represented by%
\begin{equation}
U_{j}(\Phi _{j})=g_{j,1}\Phi _{j}\!+\!\frac{g_{j,2}\Phi _{j}^{2}}{2!}\!+\!%
\frac{g_{j,3}\Phi _{j}^{3}}{3!}\!+\!\frac{g_{j,4}\Phi _{j}^{4}}{4!}-U(\Phi
_{j}^{v}).  \label{3.11}
\end{equation}
Thus, by using (\ref{3.4}) it forthwith yields that 
\begin{eqnarray}
&&\left. \omega _{j}^{2}=\frac{D}{R_{j}^{2}}\frac{q_{j}B_{j}}{A_{j}} \!+\!%
\frac{1}{2A_{j}}\!\left[ U^{(2)}\!(\Phi _{j}^{v})\!+\!\left(\frac{2}{3}%
\right)^{D/2}\!\!\!\!G_{j,0}U^{(3)}(\Phi _{j}^{v})\right. \right.  \notag \\
&&  \notag \\
&&\left. \qquad+\frac{1}{2^{D/2 +1 }}{G_{j,0}^{2}U^{(4)}(\Phi _{j}^{v})}%
\right] .
\end{eqnarray}

Again, the condition for the existence of oscillating lumps is described by $%
\omega _{j}^{2}<0$. However, the results depend upon the sign of $%
U^{(4)}(\Phi _{j}^{v})=g_{j,4}$, such that two conditions $g_{j,4}\gtrless 0$
should be analyzed. Firstly, let us analyze the case where $g_{j,4}>0$. In
addition let us assume $A_{j}>0$ and $B_{j}>0$ and that%
\begin{equation}
\omega _{j}^{2}:=\Omega (G_{0,j}),
\end{equation}
which is a parabola with positive concavity with a minimum localized at%
\begin{equation}
G_{0,j}^{\min }=-\left( \frac{4}{3}\right) ^{D/2}\frac{U^{(3)}(\Phi _{j}^{v})%
}{U^{(4)}(\Phi _{j}^{v})}.
\end{equation}
Consequently, it yields 
\begin{eqnarray}
\Omega (G_{0,j}^{\min }) &=&\frac{D}{R_{j}^{2}}\left( \frac{q_{j}B_{j}}{A_{j}%
}\right) +\frac{1}{2A_{j}}\left\{ U^{(2)}(\Phi _{j}^{v})\right.  \notag \\
&&  \notag \\
&&\left. -2^{(D-2)/2}\left( \frac{2}{3}\right) ^{D}\frac{\left[ U^{(3)}(\Phi
_{j}^{v})\right] ^{2}}{U^{(4)}(\Phi _{j}^{v})}\right\} .  \label{4.33}
\end{eqnarray}
From the inequality $\omega _{j}^{2}<0$, we conclude that%
\begin{equation}
R_{j}\geq \sqrt{\frac{2Dq_{j}B_{j}}{\frac{1}{2}\left( \frac{2^{3/2}}{3}%
\right) ^{D}\frac{\left[ U^{(3)}(\Phi _{j}^{v})\right] ^{2}}{U^{(4)}(\Phi
_{j}^{v})}-U^{(2)}(\Phi _{j}^{v}) }}.  \label{4.34}
\end{equation}
In the above expression the denominator must be positive. As a consequence,
we immediately obtain%
\begin{equation}
D\leq \frac{\ln \left\{ \frac{2U^{(2)}(\Phi _{j}^{v})U^{(4)}(\Phi _{j}^{v})}{%
\left[ U^{(3)}(\Phi _{j}^{v})\right] ^{2}}\right\} }{\ln (2^{3/2}/3)}.
\label{4.35}
\end{equation}
From the condition (\ref{4.35}), the dimension for the existence of
oscillating lumps is realized to have an upper limit. On the other hand, the
condition $2^{3/2}/3<1$, implies that the potential must obey the constraint 
\begin{equation}
\frac{2U^{(2)}(\Phi _{j}^{v})U^{(4)}(\Phi _{j}^{v})}{\left[ U^{(3)}(\Phi
_{j}^{v})\right] ^{2}}<1  \label{4.36}
\end{equation}%
\noindent as well.

\section{An application}

In this section, in order to apply the approach presented in the previous
sections for a realistic case, let us consider the symmetric double-well
potential, which is the most important model to find both topological
defects and a wide class of problems involving phase transitions. Here, we
choose the potential to have the form%
\begin{equation}
U_{j}(\Phi _{j})=\frac{\lambda _{j}}{4}\left[ \Phi _{j}^{2}-\left( \Phi
_{j}^{v}\right) ^{2}\right] ^{2},  \label{3.13}
\end{equation}
\noindent where in Eq. (\ref{1.9}) $N=4$, $g_{j,1}=g_{j,3}=0 $, $%
g_{j,2}=-\lambda _{j}(\Phi _{j}^{v})^{2}$ and $g_{j,4}=6\lambda _{j}$. The
potential (\ref{3.13}) straightforwardly implies that 
\begin{eqnarray}
&&\left. U_{j}^{(1)}(\Phi _{j}^{v})=0\,,\qquad U_{j}^{(2)}(\Phi
_{j})=2\lambda _{j}\left( \Phi _{j}^{v}\right) ^{2}\text{,}\right. \\
&&\text{ }U_{j}^{(3)}(\Phi _{j})=6\lambda _{j}\Phi _{j}^{v}\,,\qquad
U_{j}^{(4)}(\Phi _{j})=6\lambda _{j}\text{.}  \notag
\end{eqnarray}
It is worth to emphasize that the condition provided by Eq. (\ref{4.36})
reduces to the value $2/3$. Moreover, for the sake of simplicity, we will
apply the scale $R_{j}=\tilde{R}_{j}/\sqrt{\lambda _{j}}\Phi _{j}$. \ Thus,
Eq. (\ref{4.34}) reads 
\begin{equation}
R_{j}\geq \sqrt{\frac{2Dq_{j}B_{j}}{3\left( \frac{2^{3/2}}{3}\right) ^{D}-2}}%
.  \label{3.14}
\end{equation}

By assuming $q_{j}=1$ it implies for $D=2$ that $R_{j}\geq \sqrt{6B_{j}}$.
On the other hand, choosing $D=3$ we obtain $R_{j}\gtrsim 2.42\sqrt{2B_{j}}$%
. Furthermore, Eq.(\ref{4.35}) imply that $D\leq 6$.

In what follows we depict the behavior of the oscillon as a function of the
coordinates $r$ and $t$ as well. It is remarkable the appearance of a kind
of double oscillon profile during some time along the evolution of the
fields configurations obtained in this work, as it can be observed in the $%
\chi $ profile plotted in the Figure 1. Moreover, since $\phi $ and $\chi $
come from combinations of $\Phi _{1}$ and $\Phi _{2}$ with different time
dependency, a clear beating behavior shows up in their profiles, as one can
see in the Figure 2. As a consequence, these objects would appear in
periodic bursts.

\vspace{0.5cm} 
\begin{figure}[h]
\includegraphics[scale=0.86]{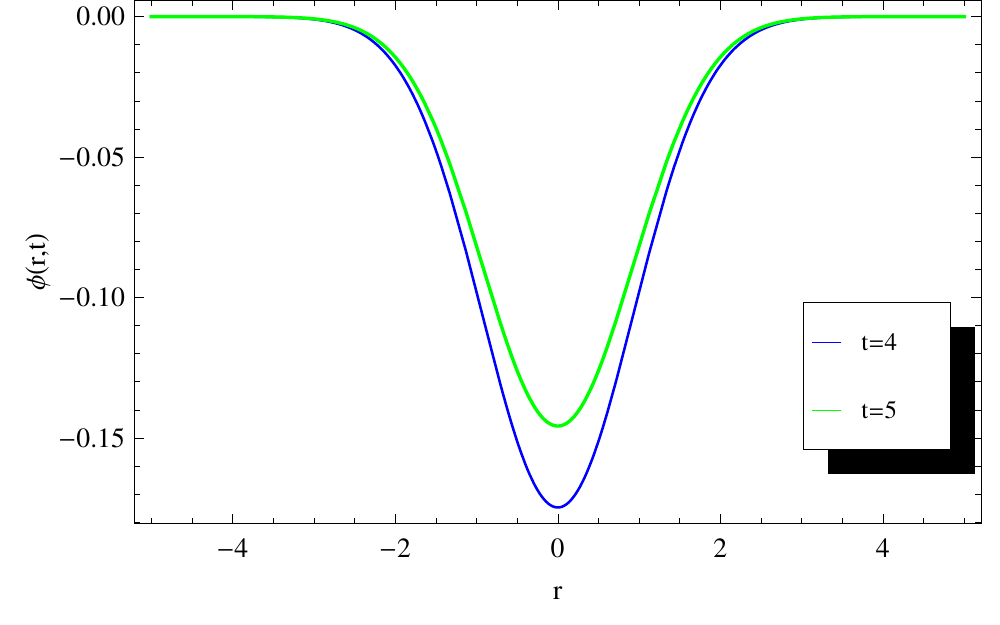} %
\includegraphics[scale=0.86]{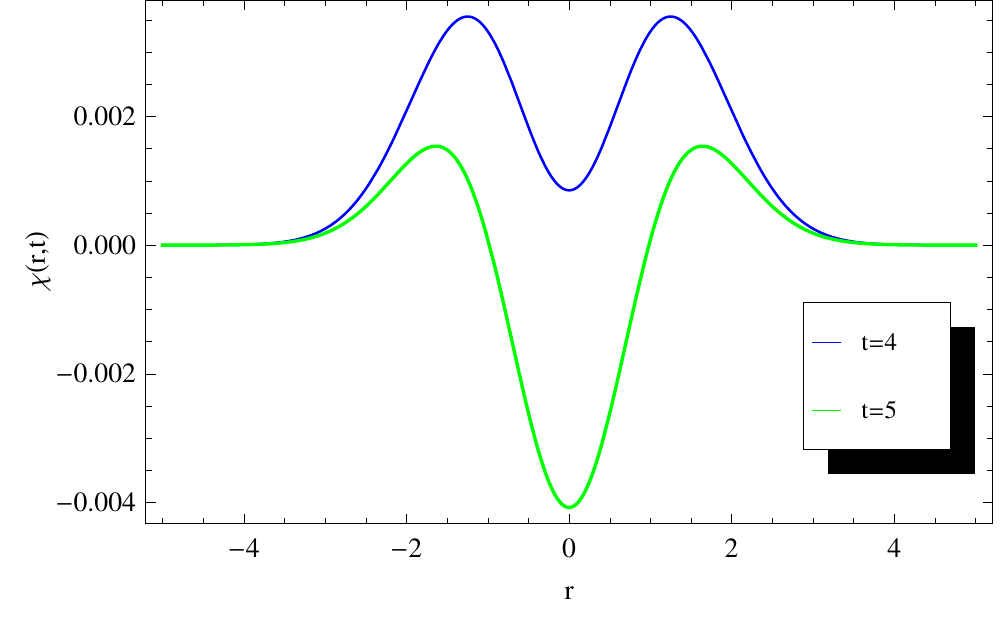}
\caption{Profile of the oscillon with $D=2$, $\protect\alpha =0.1$ and $%
\protect\lambda _{1}=\protect\lambda _{2}=1$.}
\end{figure}
\vspace{0cm} 
\begin{figure}[h]
\includegraphics[scale=0.92]{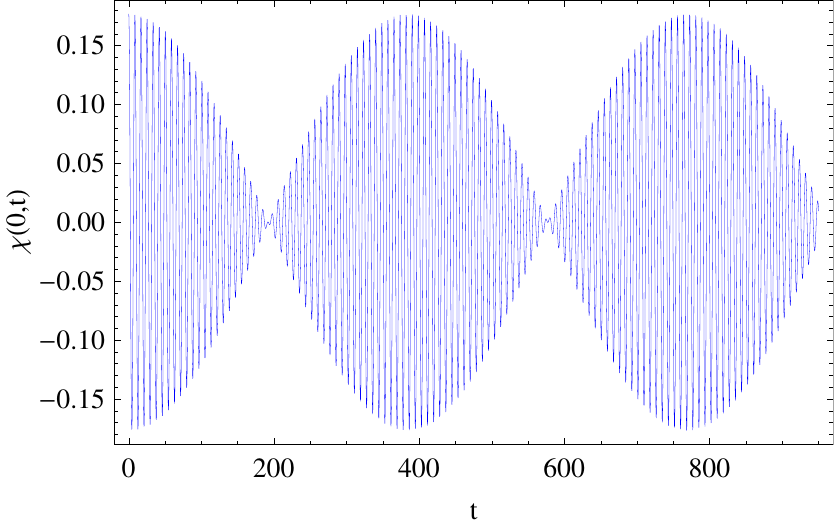} %
\includegraphics[scale=0.92]{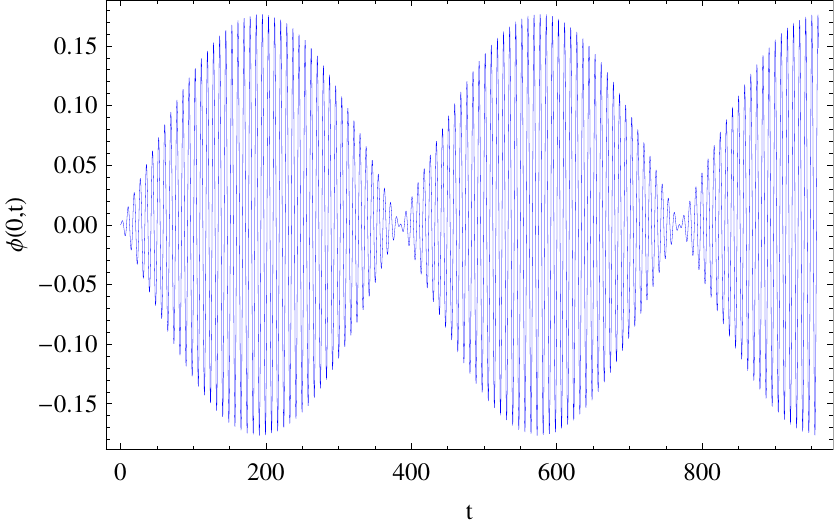}
\caption{Configurations $\protect\phi $ and $\protect\chi $ in $r=0$ for $%
D=2 $, $\protect\alpha =0.1$ and $\protect\lambda _{1}=\protect\lambda %
_{2}=1 $.}
\end{figure}
\vspace{0.5cm} 
\begin{table}[h]
\begin{center}
\begin{tabular}{||r||r||r||r||}
\hline\hline
$D$ & $\alpha $ & $R_{min}^{(1)}=\sqrt{\frac{2DB_{1}}{3\left( \frac{2^{3/2}}{%
3}\right) ^{D}-2}}$ & $R_{min}^{(2)}=\sqrt{\frac{2DB_{2}}{3\left( \frac{%
2^{3/2}}{3}\right) ^{D}-2}}$ \\ \hline\hline
$2$ & $0$ & $1.73205$ & $1.73205$ \\ \hline\hline
$2$ & $0.01$ & $1.72337$ & $1.74069$ \\ \hline\hline
$2$ & $0.03$ & $1.70587$ & $1.75784$ \\ \hline\hline
$3$ & $0$ & $2.41553$ & $2.41553$ \\ \hline\hline
$3$ & $0.01$ & $2.40342$ & $2.42758$ \\ \hline\hline
$3$ & $0.05$ & $2.34194$ & $2.48694$ \\ \hline\hline
\end{tabular}%
\end{center}
\par
\renewcommand{\tablename}{Table}
\caption{Typical size $R_{\min }$ for the symmetric double-well potential.}
\end{table}

\section{Conclusions}

In this work we investigate the consequences of the Lorentz symmetry
violation on extremely long-living, time-dependent, and spatially localized
field configurations which are called oscillons. This is accomplished in ($%
D+1$) dimensions for two interacting scalar field theories in the so-called
Standard Model-Extension context. We show that $D$-dimensional scalar field
lumps can be found in typical size $R_{\min }\ll R_{KK}$, where $R_{KK}$ is
the associated length scale of the extra dimensions in Kaluza-Klein
theories. In fact, if the fundamental gravity scale is denoted by $M$, the
length scale of the extra dimensions is $R_{KK} \sim M^{-1}(M_{\mathrm{Pl}%
}/M)^{2/(D-3)}$, and $D-3\geq 1$ is the number of extra dimensions. Thus, if 
$M \approx 1$ TeV then $R_{KK}\approx 10^{32/(d-3)} \times 10^{-19} m$ (see,
e. g., \cite{Maartens:2010ar} for a comprehensive review). Here $R_{\min }$
is shown to strongly depend on the terms that regulate the Lorentz violation
in the theory, implying either contraction or dilation of $R_{\min }$,
accordingly. Oscillons in a LSV framework present thus a set of
dimensionally-dependent properties. Moreover, the minimum radius that allows
the initial configurations to be led to oscillons is also ruled by the
dimensionality of space. If such configurations were to be probed by
observations, their sizes and energies would uniquely provide the space
dimensions. Alternatively, it can also probe the existence of $D$%
-dimensional oscillons at the TeV energy scale. In a cosmological scenario
with Lorentz symmetry breaking, we argue that in the early Universe with an
extremely high energy density and a strong Lorentz violation, the typical
size $R_{\min }$ was found highly dilated. With the Universe expansion and
cooling, a phase transition towards a Lorentz symmetry had occurred, and the
size $R_{\min }$ tended to shrink.

\bigskip

\begin{acknowledgments}
RACC thanks to UFABC and CAPES for financial support. RdR thanks to CNPq
grants No. 303027/2012-6 and No. 473326/2013-2 for partial financial
support. ASD was supported in part by the CNPq.
\end{acknowledgments}

\end{document}